\documentclass[conference]{IEEEtran}
\IEEEoverridecommandlockouts
\usepackage{graphicx} % Required for inserting images
\usepackage{xurl}
\usepackage{amsmath}
\usepackage{hyperref}
\usepackage{tabularx}
\usepackage{multirow}
\usepackage{caption}
\usepackage[most]{tcolorbox}
\usepackage{tikz, adjustbox}
\usepackage{flushend}
\newtcolorbox{answer}[1]{colback=black!5!white,
colframe=black!75!black,fonttitle=\bfseries,boxsep=2pt,left=2pt,right=2pt,top=3pt,bottom=3pt,
title={#1}\centering}

\begin{document}
\title{Deciphering WONTFIX: A Mixed-Method Study on Why GitHub Issues Get Rejected}
\author{\IEEEauthorblockN{Anonymous Authors}}
\author{\IEEEauthorblockN{J. Alexander Curtis}
\IEEEauthorblockA{\textit{Department of Computer Science} \\
\textit{Boise State University}\\
Boise, ID. USA \\
alexcurtis@u.boisestate.edu}
\and
\IEEEauthorblockN{Sharadha Kasiviswanathan}
\IEEEauthorblockA{\textit{Department of Computer Science} \\
\textit{Boise State University}\\
Boise, ID, USA \\
sharadhakasivisw@u.boisestate.edu}
\and
\IEEEauthorblockN{Nasir U. Eisty}
\IEEEauthorblockA{\textit{Department of EECS} \\
\textit{The University of Tennessee}\\
Knoxville, TN, USA \\
neisty@utk.edu}
}

\date{April 2025}
\maketitle

\begin{abstract}
\textit{Context}: 
The ``wontfix'' label is a widely used yet narrowly understood tool in GitHub repositories, indicating that an issue will not be pursued further. Despite its prevalence, the impact of this label on project management and community dynamics within open-source software development is not clearly defined.
\textit{Objective}: 
This study examines the prevalence and reasons behind issues being labeled as \textit{wontfix} across various open-source repositories on GitHub. 
%By analyzing a large dataset of labeled issues, we gain insight into the common themes and factors influencing the decision to categorize issues as \textit{wontfix}.
\textit{Method}: 
Employing a mixed-method approach, we analyze both quantitative data to assess the prevalence of the \textit{wontfix} label and qualitative data to explore the reasoning that it was used. Data were collected from 3,132 of GitHub's most-popular repositories. Later, we employ open coding and thematic analysis to categorize the reasons behind \textit{wontfix} labels, providing a structured understanding of the issue management landscape. 
%Finally using this data to create a fine-tuned AI model to effectively identify potential wontfix issues earlier in the code maintenance process.
\textit{Results}: 
Our findings show that about 30\% of projects on GitHub apply the \textit{wontfix} label to some issues. These issues most often occur on user-submitted issues for bug reports and feature requests. The study identified eight common themes behind labeling issues as \textit{wontfix}, ranging from user-specific control factors to maintainer-specific decisions. 
%The custom AI model provides an effective solution to identify potential wontfix issues earlier in the process in order to help issue creators and project maintainers, saving time and ensuring adequate assistance. These findings shed light on the complexities and nuances of issue resolution in open-source projects.
\textit{Conclusions}: 
The \textit{wontfix} label is a critical tool for managing resources and guiding contributor efforts in GitHub projects. However, it can also discourage community involvement and obscure the transparency of project management. 
%The prevalence of \textit{wontfix} labels underscores the importance of effective issue prioritization and management strategies. 
Understanding these reasons aids project managers in making informed decisions and fostering efficient collaboration within open-source communities.

\end{abstract}


\section{Introduction}\label{sec:introduction}
% Introduction Section
GitHub, an integral platform for collaborative software development, issue tracking, and project management, organizes issues through the use of predefined labels such as bug, enhancement, and \textit{wontfix}~\cite{Wang2022-labelpredictions,inproceedings}. The \textit{wontfix} label, which signifies a decision not to pursue an issue further, is often misunderstood and can significantly impact project management and community dynamics~\cite{Wang2020-whyismybugwontfix}. When a developer creates an issue on GitHub that is labeled \textit{wontfix}, they can feel rejected, upset, and demotivated~\cite{Ye2003-rt}. Understanding the rationale and implications of the usage of this label is essential for effective project governance~\cite{Zhou2012-sj}.

\textbf{Our study focuses on the prevalence and ramifications of the \textit{wontfix} label in open-source GitHub repositories.} We aim to quantify its usage, uncover common patterns among the issues to which it is applied, and explore its impact on project outcomes and community engagement. Through a detailed analysis of issue descriptions, comments and project documentation, we categorized and analyzed the recurring themes and justifications provided by the project maintainers on \textit{wontfix} issues. By understanding the underlying reasons behind the application of the \textit{wontfix} label, we have uncovered patterns and trends in issue management practices across different GitHub repositories.

We determined that around 30\% of the most popular projects on GitHub actively use the \textit{wontfix} label to organize their issues and pull-requests. These issues are most commonly paired with \textit{bug}, \textit{questions}, and \textit{enhancement} labels, providing insight into the type of issues that are often rejected. When we analyzed why this happens, through a statistically significant qualitative study, we found that the most common reasons relate to a lack of healthy discussion, user-specific environment discrepancies, and pre-existing workarounds. Surprisingly, we also discovered that non-English posts are almost certainly rejected in popular repositories, leading to questions about inclusivity in a worldwide open-source community~\cite {Ducheneaut2005-ei}.

We seek to clarify the contexts and consequences of the \textit{wontfix} label's application. By researching the types of issues most commonly rejected with this label, we can provide actionable insights for developers to enhance their submissions, thus fostering more effective issue resolution and community collaboration. Additionally, our findings will offer valuable guidance for both open and closed-source project management, potentially influencing tooling enhancements to better identify and address \textit{wontfix} issues~\cite{Guo2010-bz}. By providing a detailed analysis of the \textit{wontfix} labels' usage, underlying reasons, and consequent effects on projects; we aim to assist maintainers in making informed decisions, thereby enhancing project health and fostering a more inclusive open-source ecosystem.

\noindent We seek to answer the following research questions:
\begin{itemize}
    \item{\textbf{RQ1: What is the prevalence of the \textit{wontfix} label in open-source repositories?}}
    \item{\textbf{RQ2: What common characteristics do issues with the \textit{wontfix} label share?}}
    \item{\textbf{RQ3: What are the principal reasons for issues being labeled as \textit{wontfix}?}}
%    \item{\textbf{RQ4:} Can we predict whether an issue is likely to be labeled as \textit{wontfix} using Artificial Intelligence?}
\end{itemize}

%We expect that the findings of this research will contribute to improving issue management practices in open source projects on GitHub.  By highlighting best practices for applying the \textit{wontfix} label and its impact on project quality and community engagement, this research can help project maintainers make more informed decisions and foster a healthier and more inclusive open source ecosystem. 

\section{Related Works}\label{sec:background}

Kim and Lee~\cite{Kim2021-na} explored the dynamics of issues with multiple labels on GitHub, examining the reasons behind their closure or interaction rates. Similarly, Cabot et al.!\cite{Cabot2015-ao} focused on issues labeled with \textit{wontfix} among other labels, analyzing their implications in issue management.
Distinctively, the study by Di Sorbo et al.~\cite{Di_Sorbo2019-ep} specifically addresses the \textit{wontfix} label, applying machine learning to predict such classifications. This contrasts with our research, which aims to understand the rationale behind applying the \textit{wontfix} label to reduce the number of unresolved issues in version control systems.

Moreover, the available literature extends to the examination of GitHub's labeling practices beyond \textit{wontfix}. Tan et al.~\cite{Tan2023-xj} explored the \textit{good first issue} label, contributing methods that we adapt for our analysis of \textit{wontfix} labels. While their research aims to facilitate new developer contributions, our focus lies in understanding the rationale and consequences of labeling issues as rejected.

The phenomenon of issue staleness, not exclusive to the \textit{wontfix} label, has been reviewed by Wessel et al.~\cite{Wessel2019-cd}. Our hypothesis aligns with their findings, suggesting that the reasons behind general issue staleness may overlap with the reasons for applying \textit{wontfix} labels.

Additionally, the study by Li et al.~\cite{Li2018-jn} examines the network of interrelated issues on GitHub, providing insights into the ecosystem of issue management.
Significant work has also been performed on GitHub mining for research purposes. Izquierdo et al.~\cite{Izquierdo2015-ww} developed a tool to query GitHub issues, which was instrumental in our compilation of our dataset. Furthermore, Ye Paing et al.~\cite{Ye_Paing_Tatiana_Castro_Velez_Raffi_T_Khatchadourian2022-fv} conducted in-depth research on analyzing issue and pull request comments, offering methodologies beneficial for GitHub dataset studies. %Kalliamvakou et al.~\cite{Kalliamvakou2014-du} provides some overarching warnings to avoid common GitHub mining mistakes.

%Research has also been conducted to identify process improvements on Github issues to increase the probability of success in getting issues resolved. One such method is the use of "Issue Report Templates", which was studied with Nikeghbal et al. to find that the use of pre-filled templates significantly improves issue completedness~\cite{Nikeghbal2023-qj}.

\section{Data Collection}\label{sec:dataset}
% Data Collection
% ========================================

% Explain how we collected the data and the decisions we made.

\begin{figure}
    \centering
    \includegraphics[width=1\linewidth]{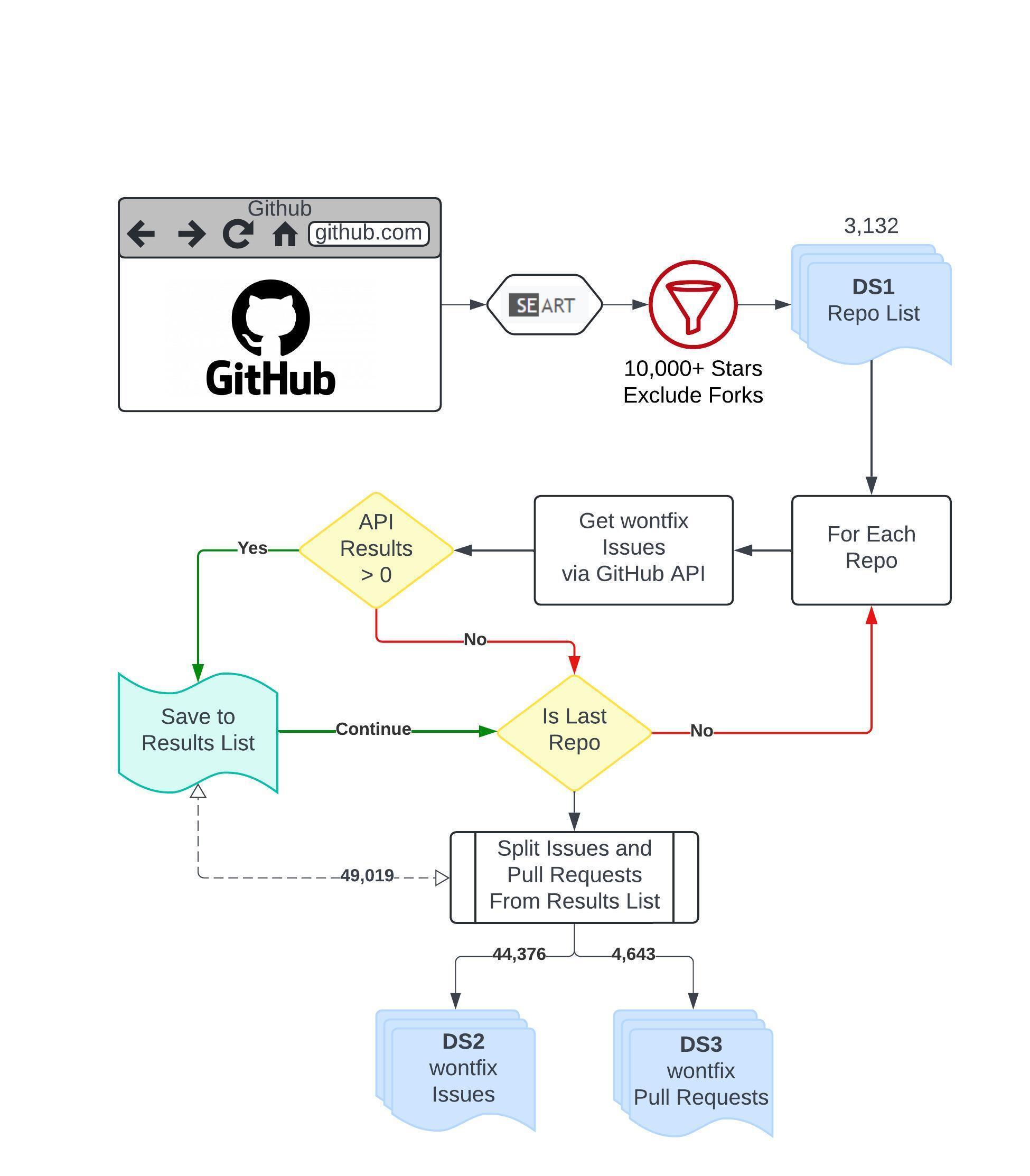}
    \caption{The data collection process pulls from the Github API to collect qualifying repositories, and then collects \textit{wontfix} issues from each repo.}
    \label{fig:data-collection-process}
\end{figure}

%Our study gathers a comprehensive dataset on the usage of the \textit{wontfix} label across the most popular GitHub repositories. 
% Initially, we intended to use GHTorrent~\cite{Ghtorrent2013} to efficiently obtain the data of issues that have the \textit{wontfix} label. Unfortunately, GHTorrent has not been updated since 2021-09-01. This makes the dataset outdated if we created it based on this data source. Therefore, we had to find other ways to collect this data and ensure that it is up to date. We will collect data from GitHub's public API, focusing on repositories that have significant community interaction across a diverse range project types, languages, and age.
% For this study, we determined that we want to understand how the \textit{wontfix} label works in ideal circumstances. Therefore, popular repositories that have large community interaction and engagement are the best sources for our data. These popular projects have enough community members to create meaningful discussions. The quantity of issues and pull requests that are created requires a sorting and organization method, which labeling provides for maintainers. Therefore, the \textit{wontfix} label becomes a useful tool in large projects and is a good place to focus our study efforts.
% \subsubsection{Collection Process}

We used the GitHub API to collect the data to search for all GitHub repositories with 10,000 or more stars as of March 23, 2024, excluding forked repositories. We decided to use GitHub stars as a proxy for popularity, as they are easy to filter by and translate equally across all project languages and project types. 
%We intentionally made the decision not to limit the search to any specific languages in order to gain a holistic understanding of how this label affects all projects. 
This search was collected and saved as Dataset 1 (referred to as \textbf{DS1}), which includes a list of qualifying repositories used in our study. The total count of these repositories was 3,132. This represents the top 0.01\% most popular projects on Github, out of 57 million public repositories~\cite{Kobayakawa2017-yb}.

Once the list of qualifying repositories is established, we have a boundary for the scope of our study. This list (DS1) will then be used as a source for collecting applicable issues and pull-requests. 
As shown in Fig.~\ref{fig:data-collection-process}, we iterate through each repository in \textbf{DS1} and retrieve all issues labeled as \textit{wontfix} for each repository. Once all \textit{wontfix} issues have been collected, we then separate the issues from the pull requests. 
%Semantically, Github characterizes every pull-request as an issue. However, within the context of a project these are displayed as separate resources and treated differently by project maintainers. Our study aims to analyze these two as distinct resource types. This approach replicates their usage on the Github website. We emphasize this distinction because their purposes and influences differ. Therefore we will take the results from collecting all issues and separate the issues (discussions, feature requests, bug reports, etc) from the pull-requests (requests to merge code).
%Our data collection process is visualized in Figure~\ref{fig:data-collection-process}. 

The final process results in two distinct datasets:

\begin{itemize}
    \item \textbf{DS1}: repositories matching study criteria
    \item \textbf{DS2}: \textit{wontfix} issues from qualifying repositories
    %\item \textbf{DS3}: \textit{wontfix} pull-requests from qualifying repositories
\end{itemize}

%Each of these datasets will be referenced throughout our study to answer different parts of the research study. These datasets are available for download and verification as referenced in Section~\ref{sec:data-availability}. The data collection scripts where written in Python and are also provided for download.

%Using the collected data, we will address the research questions via a mixed-methods approach, combining quantitative and qualitative analyses. This hybrid approach will allow an exploration of both the statistical prevalence and contextual implications of the \textit{wontfix} label in open-source projects.

\section{RQ1: What is the prevalence of the \textit{wontfix} label in open-source repositories?}\label{sec:rq1}
% RQ1: How Common is the \textit{wontfix} Label in Open Source Repositories?
% =======================================================================================
% Start by explaining the motivation and goal or value of getting this question answered.

% In the dynamic landscape of open-source software development, the management of issues and feature requests is pivotal for effective collaboration and project maintenance. A particular aspect that has garnered attention is the use of the \textit{wontfix} label within these repositories. The \textit{wontfix} label, when applied, indicates that despite acknowledgment, an issue or request will not be addressed. Understanding the prevalence of this label provides essential quantitative insights into its' impact across various open-source projects. By quantifying how frequently \textit{wontfix} is employed, we can gauge the extent to which projects reject certain community-reported issues, thereby setting the stage for deeper analysis into the decision-making processes underlying this practice. 

% Ultimately, this research question aims to guide project managers and contributors in refining their management approaches to foster more efficient and collaborative open-source environments.

\begin{figure}
    \centering
    \includegraphics[width=1.2\linewidth]{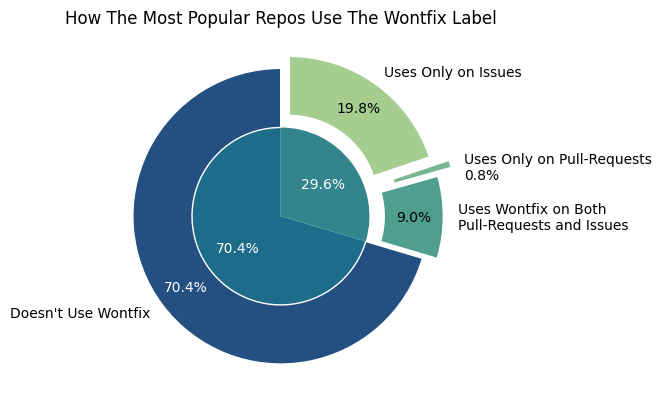}
    \caption{Analysis of Popular Repositories using \textit{wontfix} label}
    \label{fig:wontfix-freq-pie-chart}
\end{figure}
\textbf{\textit{RQ1 Methodology.}}
To investigate our research question, we begin by analyzing \textbf{DS1}, a dataset of popular GitHub repositories spanning various programming languages. First, we establish a baseline by counting the total number of repositories. We then identify those that have applied the \textit{wontfix} label to at least one issue or pull request, allowing us to measure its overall adoption.
Next, we break down label usage by distinguishing between its application to issues, pull requests, or both. This segmented analysis offers insights into how \textit{wontfix} is used across different contribution types.
Although GitHub includes the \textit{wontfix} label by default in new repositories, its use is optional. Therefore, we consider a repository as using the label only if it has actively applied it to an issue or pull request.

% Results
\textit{\textbf{RQ1 Results.}}
An analysis of 3,132 repositories found that 29.6\% use the \textit{wontfix} label on issues, pull requests, or both, as shown in the inner ring of Fig.~\ref{fig:wontfix-freq-pie-chart}. This indicates a notable, though not universal, adoption of the label.

\begin{figure}
    \centering
    \includegraphics[width=0.8\linewidth]{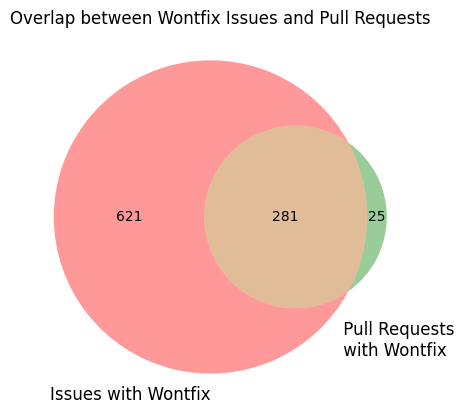}
    \caption{Venn-Diagram featuring \textit{wontfix} Issues and PRs}
    \label{fig:wontfix-frequency-venn-diagram}
\end{figure}

Of these, 902 repositories applied \textit{wontfix} to issues, while only 306 used it on pull requests. As illustrated in Fig.~\ref{fig:wontfix-frequency-venn-diagram}, most repositories are only using the \textit{wontfix} label on issues, contrasting with only 25 using it for pull requests exclusively. A total of 281 repositories applied the label to both types. The outer ring of Fig.~\ref{fig:wontfix-freq-pie-chart} shows these usage patterns as percentages of the full dataset.

The label's presence in nearly a third of repositories highlights its prominent role in issue management. 
% Maintainers use \textit{wontfix} to indicate issues outside the project's scope. 
Its higher usage on issues suggests differing strategies for handling issues versus pull requests.

% FUTURE ADDITIONS:
% 
% - Repos with highest utilization of wontfix
% - Find the average percentage of issues (wontfix issues / total issues) for the projects that use wontfix
% - 

% ANSWER BOX
\begin{answer}{Summary for RQ1}
Approximately 30\% of analyzed repositories use the \textit{wontfix} label, with a significantly higher frequency in issues compared to pull requests, highlighting its importance in issue management strategies.
\end{answer}
\vspace{4pt}

\section{RQ2: What common characteristics do issues with the \textit{wontfix} label share?}\label{sec:rq2}

\textit{\textbf{RQ2 Methodology.}}
To proceed with this research question, we gathered data on issue labels from GitHub repositories, including \textit{wontfix} labels (\textbf{DS2}) and other available types by employing a data-driven approach to investigate the similarities between issues and pull-requests marked with the \textit{wontfix} label. The data pre-processing step involved typecasting the labels column to an array and counting label occurrences. We then conducted Exploratory Data Analysis (EDA) using Python pandas to examine the label frequencies that co-occur with the \textit{wontfix} label. This analysis allowed us to identify patterns and trends leading to \textit{wontfix} categorization, contributing to a more comprehensive view of issue management strategies in open-source repositories. 

%Additionally, we utilized matplotlib and seaborn libraries for visualizations, creating bar chart plots. A word cloud was generated to highlight the most common labels and their relative importance in the context of \textit{wontfix} label frequency categorization, complementing our quantitative analysis.

% Results
\textit{\textbf{RQ2 Results.}}
The predictive analysis of label co-frequency patterns revealed insightful patterns in issues and pull requests marked with the \textit{wontfix} label. The top three labels co-occurring with \textit{wontfix} issues and pull requests were \textit{bug}, \textit{enhancement}, and \textit{question} labels, respectively. 

\begin{figure}
    \centering
    \includegraphics[width=0.8\linewidth]{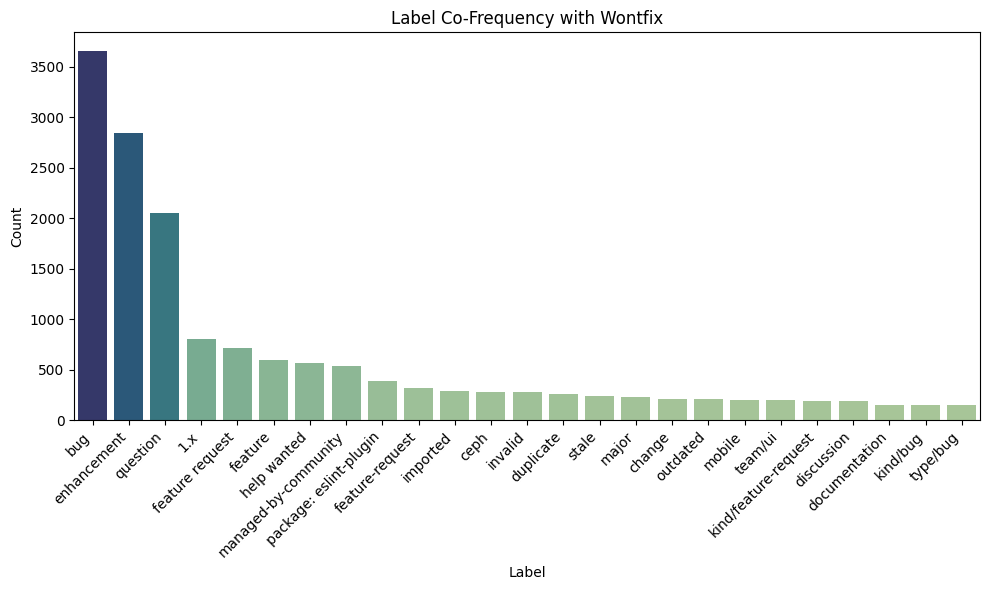}
    \caption{The frequency of other labels being paired with a wontfix issue. Counts include the total number of wontfix issues in our dataset with this label also present}
    \label{fig:label-co-frequency}
\end{figure}

The distribution of these top labels and their frequencies can be observed in Fig.~\ref{fig:label-co-frequency} plot. This observation suggests that issues and pull requests related to bugs, enhancements, and questions are more likely to be labeled as \textit{wontfix}. 
%Thus, it aids in understanding the patterns and trends leading to \textit{wontfix} categorization, contributing to a more comprehensive view of issue management strategies in open-source repositories. 

\begin{table}[h!]
\centering
\renewcommand{\arraystretch}{1.5}
\begin{tabular}{|l|r|}
\hline
\textbf{Statistic} & \textbf{Value} \\ \hline
Count & 2268 \\ \hline
Mean & 13 \\ \hline
Standard Deviation & 113 \\ \hline
Minimum & 1 \\ \hline
25th Percentile & 1 \\ \hline
50th Percentile & 2 \\ \hline
75th Percentile & 5 \\ \hline
Maximum & 3659 \\ \hline
\end{tabular}
\caption{Summary Statistics of Label Co-Frequencies}
\label{table:statistics}
\end{table}

Furthermore, the summary statistics of the data are provided in Table~\ref{table:statistics}, which offers a concise overview of the distribution and central tendency of the dataset. Most notably, it demonstrates the long tail of labels used alongside \textit{wontfix} labels. Despite returning a result of 2,268 labels, we determined that only the top 83 were statistically significant.

%The wordcloud visualization from Figure \ref{fig:wordcloud} provides a clear representation of the frequency of the correlated labels, highlighting the most common labels associated with \textit{wontfix} issues and pull-requests and might be useful for project maintainers in managing and resolving issues.

% \begin{figure}
%     \centering
%     \includegraphics[width=1\linewidth]{assets/wordcloud.png}
%     \caption{Word-cloud representing common labels found in combination with the wontfix label. Larger words represent more common associations}
%     \label{fig:wordcloud}
% \end{figure}

% ANSWER BOX
\begin{answer}{Summary for RQ2}
Commonly labeled \textit{wontfix} issues and pull requests are often associated with \textit{bug}, \textit{enhancement}, and \textit{question} labels, hinting at prevalent traits that contribute to the categorization of \textit{wontfix}.
\end{answer}
\vspace{4pt}

\section{RQ3: What are the principal reasons for issues being labeled as \textit{wontfix}?}\label{sec:rq3}
% RQ3: What common reasons are there for an issue getting labeled as \textit{wontfix}?
% =======================================================================================
% Start by explaining the motivation and goal or value of getting this question answered.

% Exploring the common reasons behind \textit{wontfix} issues, offers a broad perspective on the decision-making dynamics within open-source projects. This research question aims to uncover the diverse factors that contribute to the categorization of issues, providing a comprehensive understanding of how projects navigate and prioritize their development tasks. The motivation behind this investigation lies in optimizing issue management strategies, fostering inclusive community participation, and promoting transparency throughout project development. 

% Examining these common reasons serves multiple purposes. First, it helps project maintainers prioritize their efforts effectively by focusing on issues that align with project goals and are feasible to address. Understanding why certain issues are labeled as \textit{wontfix} also helps to refine communication and collaboration between project contributors and users. In addition, this analysis contributes to the development of best practices and guidelines for issue resolution, enhancing the overall efficiency and success of open-source initiatives~\cite{Ducheneaut2005-ei}.

\begin{figure}
    \centering
    \includegraphics[width=0.9\linewidth]{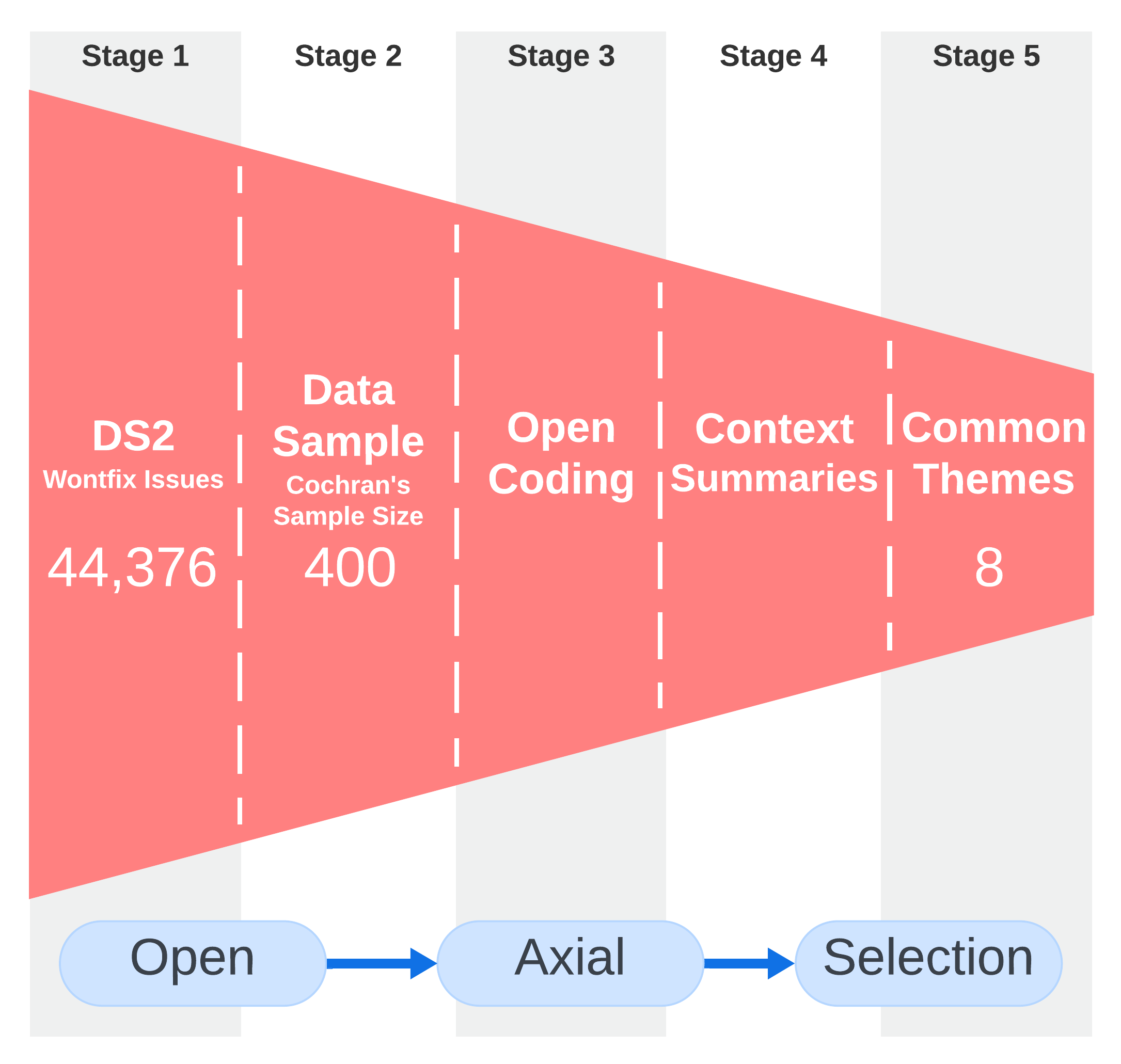}
    \caption{An overview of our grounded theory approach for selecting common themes from \textit{wontfix} issues}
    \label{fig:grounded-theory}
\end{figure}

\begin{figure}
    \centering
    \includegraphics[width=1\linewidth]{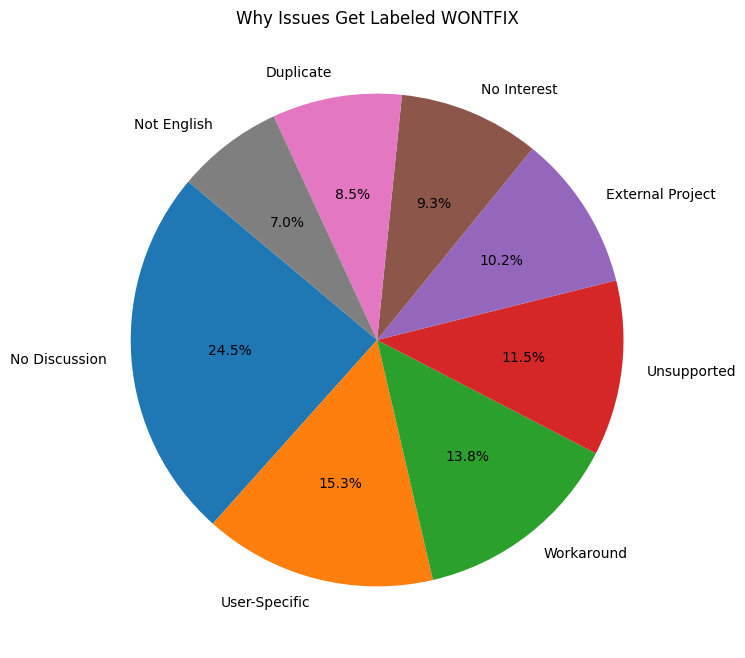}
    \caption{Common reasons why issues get labeled as \textit{wontfix}}
    \label{fig:qualitative-pie}
\end{figure}

% Methodology]
\textit{\textbf{RQ3 Methodology.}}
Our final Research Question aimed to conduct a qualitative study using a grounded theory approach~\cite{williams2019art} to understand the common reasons why issues are labeled as \textit{wontfix} in open-source repositories on GitHub. An overview of our grounded theory approach is presented in Fig.~\ref{fig:grounded-theory}. 
Our methodology involved several key steps:

\begin{itemize}
\item \textbf{Dataset Selection:} We start with a dataset containing a large number of issues from various GitHub repositories, including those labeled as \textit{wontfix}. The Grounded Theory approach adopted in our study involved this initial dataset of 44,376 total issues from \textbf{DS2}. 

\item \textbf{Sample Size Determination:} Using Cochran’s Sample Size Formula~\cite{Cochrans-2001} with a 95\%  confidence interval ($\alpha = 0.05$), we determined a minimum sample size of 382. We rounded this up to 400 for practical purposes to form a basis for our analysis. We opted to use a statistically significant sample due to the impracticality of manually analyzing all 44,376 issues, ensuring a feasible approach.

\item \textbf{Random Sampling:} From \textbf{DS2}, we randomly sampled 400 \textit{wontfix} issues to ensure a representative selection.

\item \textbf{Open Coding:} In the open coding phase, we independently read each of the 400 sampled issues in full by manually analyzing the titles and descriptions of reported issues when submitted~\cite{Di_Sorbo2019-ep}. We read a total of 1611 comments from 400 sampled issues, and then we wrote short (2-5 word) explanations for why each issue was marked as \textit{wontfix} based on the full context of comments and user behavior. This phase aimed to capture a broad range of reasons and themes.

\item \textbf{Axial Coding:} After the open coding phase, we met to compare and discuss our coding results. We identified common themes and patterns among the issues marked as \textit{wontfix}. This process of theme identification through the axial coding phase involved grouping similar reasons and categorizing them into themes.

\item \textbf{Selection Coding:} Finally, in the selection coding phase for theme validation, we refined and validated the themes and codes were assigned, thereby refined these themes as needed with the results generated from the axial coding stage. Any disagreements were resolved through discussion and consensus. After thorough analysis and discussions, we identified and agreed on eight taxonomies that were highly relevant and inclusive for understanding the reasons for issues being labeled as \textit{wontfix}. In addition, these taxonomies were broadly grouped into two categories based on the perspective of which party was most responsible for the factors that resulted in the issue being labeled as \textit{wontfix}, shown on Table \ref{table:qualitative-label}.

\item \textbf{Visualization:} The identified themes and their distribution is visualized using the pie chart in Fig.~\ref{fig:qualitative-pie} to provide a clear and structured presentation of the common reasons for issues being labeled as \textit{wontfix}.
\end{itemize}

% Results
\textit{\textbf{RQ3 Results.}}
Our manual analysis reveals several common reasons why issues are labeled \textit{wontfix}.  
%This information can be useful for project maintainers to understand why certain issues are labeled as \textit{wontfix} and how to manage them effectively. 
We broadly grouped our analysis into two scopes spanning the eight identified common themes accordingly:

\begin{enumerate}
    \item\textbf{Submitter Specific Control (65\%):} This category encompasses issues that are primarily under the control of the issue submitter. The reason for grouping these categories under Submitter Specific Control is that they all represent aspects where the issue submitter has direct influence or control over the resolution process. Whether it is about providing sufficient information, addressing user-specific issues, recognizing duplicates~\cite{Sun2011-qy}, ensuring language compatibility, or understanding external dependencies, these categories highlight scenarios where the submitter's actions or circumstances significantly impact the decision to ultimately label an issue as \textit{wontfix}. Within this category, the following 5 distinct taxonomies were identified, with approximate percentages listed below:
        \begin{enumerate}
            \item \textbf{No Discussion (25\%):} These are issues with a lack of active engagement or discussion from the submitter, contributors, or the community. This often occurs when the original submission is too detailed, complex, or vague, leading to a lack of meaningful interaction.
            \item \textbf{User-Specific (15\%):} This category comprises issues unique to the user's environment, such as specific machine or hardware settings. These issues often involve errors or problems that cannot be replicated outside of that particular environment.
            \item \textbf{External Project (10\%):} These issues arise from external dependencies that the project cannot directly update or manage. They are typically out of scope for the current project and may belong to a different project or team.
            \item \textbf{Duplicate (8\%):} Issues in this category are already addressed in later project versions or are currently in progress with a fix or feature that would resolve the concern without additional contributions. Another issue or solution covers the same issue, making further action unnecessary.
            \item \textbf{Not English (7\%):} These are discussions or submissions in languages other than English. Due to accessibility or language barriers, such issues are often closed immediately.
        \end{enumerate}

    \item\textbf{Maintainer Specific Control (35\%):} This category includes issues where control over resolution lies primarily with the project maintainers. The main reason for this grouping is that it reflects the nature of the issues and the decision-making authority with respect to their resolution. We identified the following taxonomies:
        \begin{enumerate}
            \item \textbf{Workaround (14\%):} Issues where a satisfactory workaround exists, negating the need for further development. This indicates that while the issue may exist, a viable workaround is already in place, reducing the urgency for direct resolution.
            \item \textbf{Unsupported (11\%):} Issues considered technically infeasible, too difficult to implement, or adding undesired complexity compared to expected value. These issues often require significant resources or changes that may not align with the project's goals or capabilities, leading to the decision not to support or address them directly.
            \item \textbf{No Interest (9\%):} Issues in which maintainers or the community show a lack of interest in addressing the problem or implementing the suggested feature. This category is significant because it highlights situations where there is a consensus or decision from the project maintainers not to pursue certain issues.
        \end{enumerate}
\end{enumerate}

These taxonomies and their estimated percentages offer a structured view of common reasons for \textit{wontfix} labels, based on control perspectives, as shown in Table~\ref{table:qualitative-label}.

% Tabular Column for Qualitative Study - Findings

\begin{table*}[]
\centering
\setlength{\tabcolsep}{10pt} % Default value: 6pt
\renewcommand{\arraystretch}{1.5}
\begin{tabularx}{\textwidth}{|l|c|X|} % Use X for the third column to manage overflow
\hline
\multicolumn{1}{|c|}{\textbf{RESPONSIBILITY}} & \textbf{CATEGORIES} & \textbf{EXPLANATION} \\ \hline
\multirow{5}{*}{Submitter specific}       & No Discussion       & No replies or discussion from submitter, contributors, or community. Often caused by: Original submission is too detailed and complex, or sometimes too vague. \\ \cline{2-3} 
                                          & User-Specific       & Unique problems within the user’s environment like machine or hardware settings. User errors that are not reproducible outside of the user’s environment. \\ \cline{2-3} 
                                          & Duplicate           & Resolved in later versions. A fix or feature is already in-progress which would resolve the submitted concern without additional project contributions. Another issue covers this same concern. \\ \cline{2-3} 
                                          & Not English         & Discussions that were not in the English language. \\ \cline{2-3} 
                                          & External project    & Submitted concern was due to an external dependency that the project is unable to update or manage. The request is out-of-scope for the current project and belongs in a different project. \\ \hline
\multirow{3}{*}{Maintainer specific}      & No interest         & Maintainers declined due to low interest. No community interest. Incompatible with the project mission. \\ \cline{2-3} 
                                          & Workaround          & A satisfying fix was provided to the submitter that resolves the initial inquiry without requiring further project development. Fixed with community help based on existing workaround. \\ \cline{2-3} 
                                          & Unsupported         & Technically infeasible. Difficult to implement. Adds undesired complexity compared to expected value. \\ \hline
\end{tabularx}

\caption{Qualitative Label Analysis}
\label{table:qualitative-label}
\end{table*}

% ANSWER BOX
\begin{answer}{Summary for RQ3}
Identification of 8 common themes behind labeling issues as \textit{wontfix} was accomplished in the study. Understanding these reasons enables project managers to significantly improve their issue prioritization and resolution strategies.
\end{answer}
\vspace{4pt}

% Pie Chart for Qualitative Study

% \section{RQ4: Can we predict whether an issue is likely to be labeled as WONTFIX using Artificial Intelligence?}\label{sec:rq4}
% \input{sec_7_RQ4}

\section{Discussion \& Findings}\label{sec:discussion}
% Discussion

We highlight key findings and their implications for maintainers, contributors, and researchers.

\subsection{\textbf{Wontfix issues are in Common Usage}}

We discovered that \textit{wontfix} labels are particularly prevalent among GitHub's most popular repositories. Of the top 3,132 repositories with the highest starred rating on GitHub, 1,173 have the \textit{wontfix} label configured, with 927 actively applying it to issues, pull requests, or both.

Notably, the \textit{bug} label most frequently accompanies the \textit{wontfix} label. This indicates that many of the issues and pull requests dismissed by maintainers are related to bugs and errors within the project. Our qualitative research uncovered various reasons why maintainers might dismiss bugs that would seem critical to address. A prevalent reason is that the bug is related to an external dependency, thereby placing the responsibility for a fix on another project~\cite{Sun2011-bq}.

GitHub underscores the importance of the \textit{wontfix} label by including it as one of the nine default labels for new repositories. Our study also revealed that projects often creatively rename this label, adding emojis and other symbols to convey its meaning, such as a skull and bones, a red ``X'', and various versions of a no-entry sign. Additionally, variations like \textit{status: Wontfix}, \textit{wont-fix}, and \textit{no-fix} are common. This diversity in labeling suggests that the actual use of the \textit{wontfix} label may be more widespread than our data initially indicates. A similar study of these personalized abstract labels corroborates the findings of Wang et al.~\cite{Wang2022-labelpredictions}.

% Our qualitative study backed up these findings with 11.5\% of issues being labeled as \textit{wontfix} because the maintainers were not interested in supporting or building out a suggestion from the community. Likewise another 9.3\% did not garner enough interest from maintainers or community members. Therefore the \textit{wontfix} label is an important indicator by maintainers to delineate the scope of a project and their desired involvement with it.

\subsection{\textbf{The Submitter's Role in Preventing \textit{wontfix} Issues}}

Our qualitative study revealed that in approximately 65\% of cases where an issue is labeled as \textit{wontfix}, the submitter could have prevented this outcome. Often, \textit{wontfix} labels are applied not because maintainers are unwilling to address the issue, but because the issue stems from circumstances such as environment-specific setups or misunderstandings about the project's scope, which the submitter could have identified beforehand.
Key factors within the submitter's control include:

\begin{itemize}
    \item Ensuring thorough discussion with adequate detail and context (24.5\%).
    \item Writing the issue in English to engage the broader GitHub community (7\%).
    \item Verifying whether the issue has already been reported or documented (8.5\%).
    \item Assessing if the problem is unique to the submitter's system (15.3\%).
    \item Determining if the issue pertains directly to the project or a dependent library (10.2\%).
\end{itemize}

These findings offer practical tips to help submitters avoid having their issues labeled \textit{wontfix}:

\begin{itemize}
    \item \textbf{Foster Meaningful Discussion:}
\textit{wontfix} labels often stem from poor issue descriptions. Submitters should provide clear, focused context to spark discussion—avoiding vague titles or overly complex posts. Break down complex topics and add details progressively~\cite{Tan2019-ka}.

\item \textbf{Use English for Broader Engagement:}
Non-English issues are frequently ignored, especially in large projects. Use translation tools to make issues accessible to the wider community.

\item \textbf{Clarify the Scope Before Submission:}
Issues tied to external dependencies are commonly labeled \textit{wontfix}. Confirm the source and contact the relevant project first, linking back if needed~\cite{Zhou2018-ka}.

\item \textbf{Check for Local Issues:}
Many problems are system-specific. Confirm your setup and include tested configurations to avoid mislabeling.
\end{itemize}

\subsection{Documentation Prevents \textit{wontfix}}

A frequent grievance among maintainers is that many issues could be prevented if users took the initiative to consult the readily available documentation. Most leading projects boast meticulously crafted documentation to address frequent and unnecessary questions or troubles. A well-known adage on GitHub, ``RTFM'', which stands for ``\textbf{R}ead \textbf{T}he \textbf{F}(explicative) \textbf{M}anual'', epitomizes this sentiment. Many issues are immediately marked as \textit{wontfix} using this or similar direct advice. It is imperative that submitters rigorously consider and read the documentation before escalating an issue.

% This dispels the notion that \textit{wontfix} labels are just passive-aggressive ways to discourage work, or to embarrass developers. In most of the cases, the submitter could identify these problems before submitting or they could put in some additional effort to improve the issue's success rate at being worked on. We submit that this finding should be encouraging to those that  

% Discuss what is in the maintainers Control

\subsection{Inclusivity and Auto-Translation Tools}

A common reason for immediate issue rejection on GitHub is the use of non-English languages. In our sample, all non-English issues received no response and were promptly marked \textit{wontfix}. Although GitHub does not enforce a language, English is the de facto standard in top repositories, limiting participation for non-English speakers~\cite{Balali2018-ia}.

This highlights a need for built-in translation tools to make GitHub more inclusive. Future research could explore whether such tools reduce \textit{wontfix} labels on non-English issues and support broader global engagement.

\section{Future Work}\label{sec:future-work}

%The research has provided insights into the prevalance and implications of using the \textit{wontfix} label in open-source repositories. 
While our findings deepen understanding of issue management, several areas warrant further study.

\textit{\textbf{Impact on Community Sentiment.}}
Future work could explore whether \textit{wontfix} labels contribute to negative interactions or lower morale by analyzing communication patterns.

\textit{\textbf{Project Health Metrics.}}
Developing metrics to assess project health, such as contributor turnover, issue resolution time, and community activity, could reveal how \textit{wontfix} usage correlates with project vitality~\cite{mackorisk}.

% \textit{\textbf{Reopened Wontfix Issues.}}
% Studying issues initially marked as \textit{wontfix} but later resolved could shed light on changing priorities or mislabeling, offering insights into project dynamics.

\textit{\textbf{Automated Labeling Tools.}}
Exploring AI/ML tools to manage or predict \textit{wontfix} usage could improve label accuracy and support healthier project workflows.

%These directions can build on our findings, advancing understanding of issue management and its impact on open-source development.

\section{Threats to Validity}\label{sec:threats}
For \textbf{internal validity}, the main concern lies in potential misclassification during data extraction from GitHub, especially with inconsistent use of the \textit{wontfix} label across projects. Interpretation of qualitative data, such as comments, may also introduce subjectivity. To address this, multiple coders reviewed the data and resolved differences through consensus.

\textbf{External validity} may be limited by our focus on active and popular repositories, which may not reflect practices in smaller or less active projects. Additionally, our findings from open-source GitHub projects may not generalize to private repositories or other platforms with different norms.

% Conclusion ================================================================================================
% \section{Conclusion}\label{sec:conclusion}
% In conclusion, our study highlights the prevalence and role of \textit{wontfix} labels, used by nearly 30\% of open-source repositories. Co-frequency analysis showed that \textit{bug}, \textit{enhancement}, and \textit{question} are the most common labels paired with \textit{wontfix}, offering insight into issue management strategies.

% Our grounded theory analysis of 400 sampled issues revealed common themes behind label usage, reflecting the collaborative decision-making within open-source communities.

% These findings enhance our understanding of issue resolution practices and offer a foundation for future research. They can help maintainers refine workflows and foster more effective, community-driven development.

% Data Availability ================================================================================================
\section{Data Availability}\label{sec:data-availability}
To facilitate replications, we provide the datasets:

\href{https://figshare.com/s/0145ced67fe3a7559646}{https://figshare.com/s/0145ced67fe3a7559646}

\bibliographystyle{ieeetr}
\bibliography{xreferences}

\end{document}